\documentclass[10pt,a4paper]{article}
\usepackage[utf8]{inputenc}
\usepackage{amsmath}
\usepackage{amsfonts}
\usepackage{amssymb}
\usepackage{caption}
\usepackage{graphicx}
\begin{document}

\centerline{\LARGE {\bf Mesoscopic system: semiclassical versus quantum}}
\centerline{ Kanchan Meena$^{1}$, P. Singha Deo$^{1}$, and A. M. Jayannavar$^{2}$}
\centerline{$^{1}$S. N. Bose National Center for Basic Sciences, Kolkata, 700106 India}
\centerline{and $^{2}$Institute of Physics, Bhubaneshwar, 751005 India}
\noindent{\it Abstract:} We explain a hierarchy of Friedel sum rule like formulas that help us understand response of mesoscopic systems
to applied electric and magnetic fields. The formulas can be derived fully quantum mechanically and then there is a mathematical way to argue a semiclassical
limit which may make the formulas practically relevant. But the mathematical prescription is not physically relevant in presence of a Fano resonance.
A proper understanding of what exactly is the physical semiclassical limit is therefor not known and require future work.

\section*{Introduction}

Mesoscopic physics has emerged as a very fascinating subject of research as it dwells in the intermediate regime of classical and quantum worlds.
The usual methods of condensed matter physics and statistical mechanics cannot be extended to these systems because the basic axioms fail.
This is essentially due to the fact that these samples are so small and can be subjected to such low temperatures that intrinsic length
scales of the material of the sample face competition from sample dimensions which can become as low as the elastic and in-elastic mean free paths
of the carriers like electrons or phonons.
We will look at a novel approach to electronic properties of a mesoscopic sample with 
applied electric and magnetic fields. Response of a sample to applied fields always depend on the quantum states involved.
We will provide a brief introduction 
to a hierarchy of density of states (DOS) that consist of local 
partial density of states, emissivity, injectivity, injectance, emittance, 
partial density of states, and finally the well known local density of 
states and the density of states and their connections to scattering phase shifts via Friedel sum rule (FSR) like formulas. 
Except for the last two there are no 
known analogues for bulk systems that are treated within the scope of 
statistical mechanics. 

\section*{The system}

In the figure 1 we show a typical mesoscopic sample which is the shaded region.
This sample is generally made up of some metal or semiconductor.
It can have several leads attached to it that are indexed $\alpha$,
$\beta$, $\gamma$, $\delta$, etc. These leads are quantum wires that inject or absorb electrons like that of a voltage probe or current probe.
In other words voltmeters or ammeters have to be modeled through such leads. Electron dynamics in the sample and the leads is fully determined by
Schrodinger equation and for an electron going from lead $\alpha$ to lead $\beta$ one has to solve the quantum mechanical scattering problem to find
the scattering amplitude $S_{\beta \alpha}$. Note that the outgoing lead index $\beta$ comes first and the incoming lead index comes next as the
scattering matrix connects outgoing wave vector to the incoming. 
These lead indices appear in all the formulas below showing the
importance of these leads in mesoscopic systems. The $\beta$th lead is
drawn in a special way signifying the tip of a scanning tunneling microscope (STM) tip which too can inject or absorb electrons. 
The contact between the STM tip and the point of
contact at $\textbf{r}$ can be controlled and this is often used as a standard experimental tool and so its role in context of a mesoscopic system
too has to be studied along with voltage and current probes.
Mesoscopic response is essentially determined by the members of this hierarchy.
Essentially material specific parameters like dielectric constant or permeability do not determine mesoscopic response as the axioms of statistical
mechanics break down for these small samples. The states involved in an experimental set up and their nature has to be determined from these
scattering matrix elements and the leads model the probes of the set up.

\begin{figure}[bt]
\centering
\includegraphics[width=.6\textwidth, keepaspectratio]{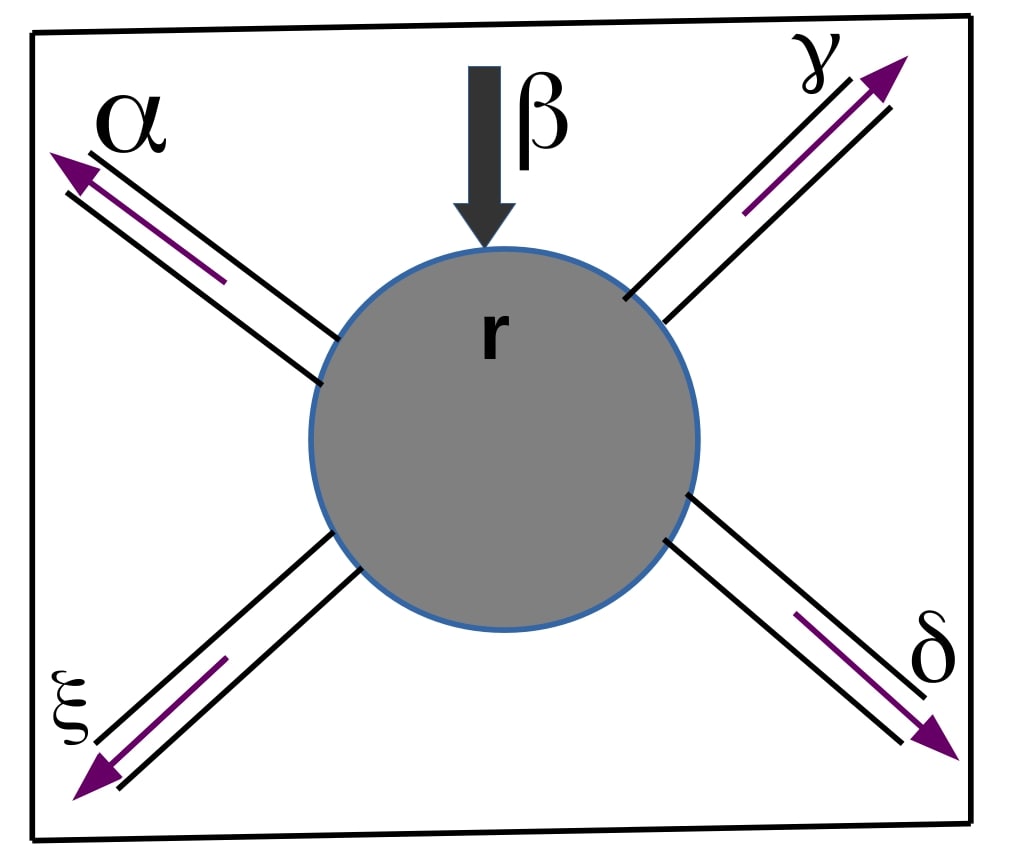}
\captionsetup{labelformat=empty}
\caption{\label{fig3}
Fig. 1 We show a mesoscopic set up where there are many leads $\alpha$, $\beta$, $\gamma$, $\delta$ etc., attached to a sample (shaded region).
The lead $\beta$ is special in the sense that it is an STM tip that can deliver (or draw) current to (or from) a particular point $\textbf{r}$ in the sample.
All other leads (minimum one other) are fixed and draw (or deliver) current from (or to) the sample.}
\end{figure}

\section*{Hierarchy of mesoscopic formulas}
We discuss a hierarchy of relations between scattering phase shifts and DOS. It starts with Larmor precession time given as \cite{but}
\begin{eqnarray}
\tau_{lpt}(E, \alpha, \textbf{r}, \gamma)=-\frac{\hbar}{4\pi i |s_{\alpha\gamma}(E)|^{2}}\left[s_{\alpha\gamma}^*\frac{\delta s_{\alpha\gamma}}
{e\delta U(\textbf{r})}- \frac{\delta s_{\alpha\gamma}^*} {e\delta U(\textbf{r})}s_{\alpha\gamma} \right]
\end{eqnarray}
A detailed derivation of this formula is given in reference \cite{but, deo}.
Here $U(\textbf{r})$ is the internal potential inside the shaded region, $e$ is the electronic charge, $E$ is the incident energy of the
electron scattered from lead $\gamma$ to lead $\alpha$, and $\frac{\delta}{\delta U(\textbf{r})}$ signifies a functional derivative with
respect to $U(\textbf{r})$.
Electrons in an ensemble
are indistinguishable fermions that occupy one state each at zero temperature below Fermi energy. Thus we define
local partial density of states (LPDOS) $\rho_{lpd}$ in the same sense as it is defined on a 1D line as $\frac{|s_{\alpha\gamma}(E)|^{2}}{\hbar} \tau_{lpt}$
\cite{but}.
\begin{eqnarray}
\rho_{lpd}(E, \alpha, \textbf{r}, \gamma)=-\frac{1}{4\pi i}\left[s_{\alpha\gamma}^*\frac{\delta s_{\alpha\gamma}}
{e\delta U(\textbf{r})}-\frac{\delta s_{\alpha\gamma}^*} {e\delta U(\textbf{r})}s_{\alpha\gamma} \right]
\end{eqnarray}
Note that $\rho_{lpd}$ is only defined between two prescribed leads which in the above formula are $\alpha$ and $\gamma$.
Eq. 2 follows from Eq. 1 from the fact that the imaginary part of the Green's function gives the average propagation time which is also $\hbar$ times
the number of intermediate states accessed during the
propagation. The propagation involves electrons that can be identified as an incoming classical particle in lead $\gamma$ as it comes from a classical
reservoir (say the terminal of a battery) and involves identical
classical electrons going to a classical reservoir via lead $\alpha$ where $\rho_{lpd}$ is a
count (or measure) 
in real number associated with the physical coordinate $\textbf{r}$ but there will be no attempt to identify or observe any electron at $\textbf{r}$ fully
satisfying the axioms of quantum mechanics. This count is called local partial density of states at the point $\textbf{r}$.
This is in a situation when the STM tip is not attached at the point $\textbf{r}$ or is attached through such a weak coupling that the states
in the system are not disturbed.
At a given incident energy $E$, electrons that are incident along input channel $\gamma$ form a quantum ensemble of identical particles. 
Every member of the ensemble is not scattered to output
channel $\alpha$. But those that are scattered to output channel $\alpha$, are done with an amplitude of $s_{\alpha \gamma}$ which means for unit incident
flux there are exactly $|s_{\alpha \gamma}|^2$ of them and some of them occupy a state (a local partial state at $\textbf{r}$). 
Doubling the input flux in $\gamma$ will double the number of electrons that are scattered to channel $\alpha$ as long as we are not in the limiting situation of a 
completely filled band. It was shown that linear superposition of incoming wave-functions cannot happen in the leads due to phase randomization
beyond the coherence length \cite{deo} but linear superposition will always happen
inside the shaded region.
If we want to consider many body effects in the leads then it again requires linear superposition of single particle states of incoming particles
to accommodate for the antisymmetric property of the many body wave-function and so the phase randomization 
will also eliminate many body effects in the leads.
However, many body effects and formation of Slater determinants is not ruled out inside the scatterer. This is consistent with observations in mesoscopic experiments
because if such Slater determinants extend beyond the sample into the leads then it will result in orthogonal catastrophe which is never observed.
This is further supported by the calculations in \cite{mukh}
which shows that Eq. 2 is also
valid in presence of electron-electron interaction inside the sample.

We get partial density of states by integrating (physically this implies averaging) $\rho_{lpd}(E, \alpha, \textbf{r}, \gamma)$ over the spatial coordinates of the sample.
\begin{eqnarray}
\rho_{pd}(E, \alpha, \gamma)=-\frac{1}{4\pi i}\int_{\Omega}d^{3}r\left[s_{\alpha\gamma}^*\frac{\delta s_{\alpha\gamma}}{e\delta U(\textbf{r})}-
\frac{\delta s_{\alpha\gamma}^*}{e\delta U(\textbf{r})}s_{\alpha\gamma} \right]
\end{eqnarray}
Here $\Omega$ stands for the spatial region of the sample that is the shaded area in figure 1.
Note that $\rho_{pd}$ is also defined between two specified leads $\alpha$ and $\gamma$.
To get injectivity $\rho_{i} $ of a specific lead $\gamma$ we sum $\rho_{lpd}$ in Eq. 2 over all the outgoing channels $\alpha$, which means 
$\alpha$ below stands for $\alpha$ and $\beta$ and $\delta$ etc.
\begin{eqnarray}
\rho_{i}(E, \textbf{r}, \gamma)=-\frac{1}{4\pi i}\sum_{\alpha}\left[s_{\alpha\gamma}^*\frac{\delta s_{\alpha\gamma}}{e\delta U(\textbf{r})}-
\frac{\delta s_{\alpha\gamma}^*} {e\delta U(\textbf{r})}s_{\alpha\gamma} \right]
\end{eqnarray}
Similarly to get emissivity $\rho_{e}$, we sum $\rho_{lpd}$ in Eq. 2 over all possible incoming channels $\gamma$.
\begin{eqnarray}
\rho_{e}(E, \alpha, \textbf{r})=-\frac{1}{4\pi i}\sum_{\gamma}\left[s_{\alpha\gamma}^*\frac{\delta s_{\alpha\gamma}}{e\delta U(\textbf{r})}-
\frac{\delta s_{\alpha\gamma}^*} {e\delta U(\textbf{r})}s_{\alpha\gamma} \right]
\end{eqnarray}
Injectivity is
like a correlation function between
two spatial points $\gamma$ and coordinate $\textbf{r}$ where the lead index $\gamma$ is actually a spatial index signifying the spatial
point where the lead is attached. Of course here there will be no ensemble averaging and yet these quantities can be
probed experimentally without involving averaging over $U(\textbf{r})$.
Hopefully, something like an STM tip can attach to a single point in its closest
approximation and can be used to change the local potential at a point infinitesimally without any tunneling between the tip and the sample. 
The tip can be also brought closer to the sample and then actual exchange of current can take place between the tip and the sample.
All such currents are determined by one or several members of the hierarchy according to the situation \cite{but, deo}.
What $\rho_i$ in Eq. 4 means is that for those
particular electrons that are coming from lead $\gamma$ irrespective of to which channel it is going, the relevant part of DOS at the point $\textbf{r}$
is $\rho_i$. All other formulas can be similarly interpreted.

From the injectivity or emissivity one can integrate over $\textbf{r}$ and arrive at higher members of the hierarchy called injectance and emittance.
Injectance can thus be defined as from Eq. 4
\begin{eqnarray}
\rho(E,\gamma)=-\frac{1}{4\pi i}\int_{\Omega}d^{3}r\sum_{\alpha}
\left[s_{\alpha\gamma}^*\frac{\delta s_{\alpha\gamma}}{e\delta U(\textbf{r})}-\frac{\delta s_{\alpha\gamma}^*}{e\delta U(\textbf{r})}
s_{\alpha\gamma} \right] 
\end{eqnarray}
Injectance can be directly measured and also alternately calculated and verified for a given $U(\textbf{r})$ as will be discussed below.
Local density of states (LDOS) can be defined by summing the RHS of Eq. 2 over $\alpha$ and $\gamma$.
\begin{eqnarray}
\rho_{ld}(E, \textbf{r})=-\frac{1}{4\pi i}\sum_{\alpha\gamma}\left[s_{\alpha\gamma}^*\frac{\delta s_{\alpha\gamma}}{e\delta U(\textbf{r})}-
\frac{\delta s_{\alpha\gamma}^*}{e\delta U(\textbf{r})}s_{\alpha\gamma} \right]
\end{eqnarray}
Integration of this over $\textbf{r}$ gives the DOS
\begin{eqnarray}
\rho_{d}(E)=-\frac{1}{4\pi i}\sum_{\alpha\gamma}\int_{sample} d^3r \left[s_{\alpha\gamma}^*\frac{\delta s_{\alpha\gamma}}{e\delta U(\textbf{r})}-
\frac{\delta s_{\alpha\gamma}^*}{e\delta U(\textbf{r})}s_{\alpha\gamma} \right]
\end{eqnarray}
\begin{eqnarray}
\rho_{d}(E)=-\frac{1}{2\pi}\sum_{\alpha\gamma}\int_{sample} d^3r \left[|s_{\alpha\gamma}|^{2}\frac{\delta \theta_{s_{\alpha\gamma}}}{e\delta U(\textbf{r})}\right]
\end{eqnarray}
Here $S_{\alpha \gamma}=|S_{\alpha \gamma}|exp[i \theta_{S_{\alpha \gamma}}]$.
This is mesoscopic Friedel sum rule that relates the functional derivative of scattering phase shift with respect to local potential to DOS. 
Normally the local potential inside a sample is not known and so these formulas in terms of the functional derivatives with respect to the local potential
is not of much use to theory or practice.
For large samples this problem can be bypassed 
by making an average over all possible variations in
$U(\textbf{r})$ but that is not allowed for mesoscopic systems.

\section*{Comment on semiclassicality}

To make these formulas more relevant to theory and practice, one may consider the following substitution
\begin{eqnarray}
\int_{global} d^3r \frac{\delta }{e\delta U(\textbf{r})}= -\frac{d}{dE}
\end{eqnarray}
On substituting this in Eq. 9 we get the FSR as given in text books \cite{zim}.
Validity of this substitution is considered to be a criterion for judging mesoscopic semiclassical regime. The criterion mentioned in text books
based on WKB regime and turning points is often not relevant for mesoscopic systems where scattering can occur due to the geometry of the sample
in absence of any potential scattering.
What this substitution means is that a global up-shift in potential by a constant amount, or a constant up-shift in local potential at all points
$\textbf{r}$ is identical to a down-shift of incident energy and vice verse. An experimentalist can easily change the incident energy
slightly by adjusting the Fermi energy of the leads.
However, in Eq. 9 we do not find a global integration but a
sample integration. 
Hence the substitution may work for a mesoscopic sample when interference effects do not dominate but is not expected to work when Fano resonance
creates a transmission zero due to destructive interference.
But counter-intuitively, it does as was shown in Ref. \cite{deo} and we will describe below.
For a mesoscopic system let us start by saying that in certain parameter regimes
\begin{eqnarray}
\int_{sample} d^3r \frac{\delta }{e\delta U(\textbf{r})}\approx -\frac{d}{dE}
\end{eqnarray}
Thus
\begin{eqnarray}
\rho_{d}(E)\approx \frac{1}{2\pi}\sum_{\alpha\beta} \left[|s_{\alpha\beta}|^{2}\frac{d \theta_{s_{\alpha\beta}}}{dE}\right]
\end{eqnarray}
It will be a surprise if we find that at the Fano resonances Eq. 12 becomes exact requiring us to re-examine what is semi-classical regime.

As we have discussed, injectance is essentially an injected current. As a general argument we know injected current is of the form
$nev$ or differential current is $\frac{dn}{dE}evdE$. Electronic charge $e$ can be set to unity and if properly normalized wave-functions are 
taken then we can also drop the $v$ factor \cite{deo} making injected current to be $\frac{dn}{dE}$ at zero temperature and 
energy $E$. So we can set up an
experimental situation
wherein the differential injected current will see only the injectance as the relevant DOS and injectance can also be theoretically
calculated for a given sample potential $U(\textbf{r})$. Namely, injectance can be determined from
internal wave-function when just the $\beta$th lead carries current into the system and all other leads take current away from the system very close to
zero temperature which can also be achieved experimentally. So as per definition of DOS
\begin{eqnarray}
\rho(E,\beta)=\int_{sample} d^3r \sum_{k_{\beta}}\vert \psi(\textbf{r},\beta)\vert ^{2}\delta(E-E_{\beta ,k_{\beta}})\\
=\int_{\Omega}d^{3}r\sum_{\alpha}-
\frac{1}{4\pi i}\left[s_{\alpha\beta}^*\frac{\delta s_{\alpha\beta}}{e\delta U(\textbf{r})}-
\frac{\delta s_{\alpha\beta}^*}{e\delta U(\textbf{r})}s_{\alpha\beta} \right] 
\end{eqnarray}
\begin{eqnarray}
\approx \frac{1}{2\pi}\sum_{\alpha} \left[|s_{\alpha\beta}|^{2}\frac{d \theta_{s_{\alpha\beta}}}{dE }\right]
\end{eqnarray}
The last step is obtained from Eq. 6 and 11. Here again $E$ is the incident energy of the electron incident along lead $\beta$ and $\psi(\textbf{r}, \beta)$
is the wave-function inside the system (shaded region in Fig. 1), $E_{\beta, k_\beta}$ being the states in the lead $\beta$ with momentum $k_\beta$. 
In 1D it was shown \cite{lea} that the correction term can be evaluated to give
\begin{eqnarray}
\rho(E,1)&=&\frac{1}{2\pi}\left[\vert r\vert^{2}\frac{d\theta_{r}}{dE}+\vert t\vert^{2}\frac{d\theta_{t}}{dE}+\frac{m_{e}\vert r\vert}{\hbar k^{2}}sin(\theta_{r}+
\theta_0)\right]
\end{eqnarray}
The first two terms are obtained if we write the sum in Eq. 15 explicitly in 1D where $S_{11}=r=|r|e^{i \theta_r}$ and $S_{21}=t=|t|e^{i \theta_t}$, and
the last phase dependent term is the correction term due to interference effects.
For a Breit-Wigner resonance or for any case other than a Fano resonance, the resonance takes place at fixed energies wherein $\theta_r$ can become $n \pi$
but nothing can be said for $\theta_0$ which is sometimes referred to as "initial phase" and is related to gauge symmetry. So Leavens ans Aers \cite{lea}
argues that one way to minimize the correction terms is to go to semiclassical regime wherein a particle is either fully transmitted or fully reflected
like a classical particle.
In quantum mechanics this happens if incident energy is much greater than the potential strength or for reflection from a hard wall.
In the former case $|r|\rightarrow 0$ and in the later case there is no interference effect at all. Another way to eradicate the interference term
is to take an ensemble average wherein $(\theta_r+\theta_0)$ is a fluctuating quantity and the correction term averages to zero. Both these situations are
termed semiclassical regimes and for them Eq 11 is a good approximation. 
It was shown in \cite{deo} that Fano resonances are very different and require a new understanding of what is a 
general semiclassical regime. Again the Fano resonances (say in 1D) occur at fixed energies wherein $(\theta_r+\theta_0)$ becomes undefined because of
a transmission 0 wherein both the real and imaginary parts of transmission amplitude become 0. 
This results
in a non-symmetry dictated node (NSDN) wherein arbitrary wave-functions can be matched across the NSDN by a discontinuous phase shift. Only $\pm \pi$
discontinuities in phase shifts are observed probably because the many body wave-function only allows that \cite{deo} but this require a better explanation.
This is discussed in detail in section 2.1 of Ref. \cite{deo}.
Which means $(\theta_r+\theta_0)$ changes discontinuously by $n\pi$ independent of $\theta_0$ and Eq. 11 becomes exact in the quantum regime of a Fano
resonance.
There is a simpler point of view that Fano resonances can be seen as an interference between two independent paths 
resulting in a zero-pole pair in the transmission amplitude. The transmission zero is due to complete destructive interference.
At the single particle level
itself $\theta_0$ is locked and the correction term depends only on $\theta_r$.
This locking mechanism has been shown in Ref. \cite{yey}.
In multichannel case the transmission zeroes soften and phase drops become gradual.
But the interesting fact is that the correction term becomes zero without any ensemble averaging even if there is partial transmission and partial reflection
at a Fano resonance. All this can be generally proved for any Fano resonance irrespective of what is the exact form of $U(\textbf{r})$ \cite{deo}.

In conclusion, mesoscopic semiclassical regime may be defined by the criterion expressed in Eq. 11 if the system is away from Fano resonance. At a Fano
resonance Eq. 11 becomes exact in spite of strong quantum effects. Hence we may have to find alternate ways of quantifying a mesoscopic semiclassical regime.

\bibliographystyle{References}

\end{document}